\documentclass[epj]{svjour}
\usepackage{epsf}
\begin{document}

\title{Semiclassical singularities from bifurcating orbits}

\author{Christopher Manderfeld\inst{1} \and Henning Schomerus\inst{2}}
\institute{Fachbereich Physik, Universit\"at--Gesamthochschule Essen,
45\,117 Essen, Germany
\and
Instituut-Lorentz, Universiteit Leiden, P.\,O.~Box 9506, 2300 RA Leiden,
The Netherlands}
\date{Received: date / Revised version: date}

\abstract{ 
We study how the singular behaviour of classical systems at
bifurcations is reflected by their quantum counterpart.
The semiclassical contributions of individual periodic orbits to trace
formulae of Gutzwiller type
are known to diverge when orbits bifurcate, a situation characteristic for
systems with a mixed phase space.
This singular behaviour is reflected by the quantum system in the semiclassical
limit,
in which the individual contributions remain valid closer to a bifurcation,
while the true collective amplitude at the bifurcation
increases with some inverse power of Planck's constant (with an exponent
depending on the type of bifurcation).
We illustrate the interplay between the two competing
limits (closeness to a bifurcation and smallness of $\hbar$) 
numerically for a generic dynamical system, the kicked top.
}
\maketitle

\section{Introduction}
The semiclassical approach
allows to obtain
spectral information about quantum systems from
properties of classical periodic orbits. 
The most famous example of this quantum-classical correspondence is
Gutzwiller's trace formula for
completely chaotic (hyperbolic) autonomous systems \cite{gutz}, which expresses
the density of states as a sum of contributions from periodic orbits.
In such systems, the Lyapunov exponents $\lambda$ of all periodic orbits
are positive, and their semiclassical amplitudes
$A\propto(\sinh\lambda/2)^{-1}$ are finite.
This is not true for systems with a mixed phase space, which accommodate
also elliptic orbits with amplitude $A\propto(\sin\omega/2)^{-1}$.
The quantity $\omega=r\omega_0$,
known as the stability angle, increases linearly under $r$ repetitions of the
orbit, and either by a suitable choice of $r$ or of an external control
parameter the amplitude $A$ can become arbitrarily large. The
contribution of an individual orbit eventually diverges when $\omega/2\pi$
is an integer.  Normal-form theory \cite{meyer} shows that this is
precisely the condition for a bifurcation, the
coalescence of two or more periodic orbits. Catastrophe theory
\cite{Poston} further reveals that the divergence
comes from an illegitimate stationary-phase approximation, and provides us
with uniform
approximations (collective contributions of the bifurcating orbits)
that regularise the singular behaviour
\cite{bif,ozorio,kus,unif123,pver4,pver3,kodim2}.
The ensuing true semiclassical amplitude 
is finite at $\hbar\neq 0$, even directly at the
bifurcation.
It is clear, however, that an individual contribution
of Gutzwiller type must be valid in the strict semiclassical limit,
as soon as one does not sit precisely on a bifurcation (fixed distance
$\varepsilon$ to a bifurcation in parameter space, $\hbar\to 0$). Analyticity
at given $\hbar$ and varying $\varepsilon$ entails then that the true
amplitude at the
bifurcation ($\varepsilon=0$)
must diverge as $\hbar\to 0$, with a
power law $\sim\hbar^{-\nu}$
as it happens to be \cite{unif123}.
Some consequences of this peculiar singularity have 
been studied recently in the context of spectral fluctuations \cite{jab}.
A complete solution of this intricate problem is emerging
\cite{Berryneu}, but it is complicated because one has to
consider also more complex bifurcations
of higher codimension which are classically non-generic, but are
nevertheless relevant in the quantum realm.

In this work we investigate the interplay of the two competing limits
$\hbar\to 0$, $\varepsilon\to 0$, focusing on periodically driven systems
with one degree of freedom.
All results are relevant for autonomous systems with two degrees of freedom
as well.
For a representative system, the kicked top, we find that bifurcations
often interfere even when they are separated in phase space.
This adds additional
complexity to the problem at hand.
We use a filtering technique to extract contributions of
bifurcating orbits and find that their amplitude
corresponds well to the theoretical predictions.

The paper is organised as follows: In Section \ref{sec2}
we describe how the exponents $\nu$ for the most commonly
encountered bifurcations are derived.
In Section \ref{sec3} we present numerical results for the
kicked top. Section \ref{sec4} contains our conclusions.

\section{Semiclassical contributions at bifurcations}
\label{sec2}

Periodically driven systems 
are stroboscopically described by a unitary Floquet operator $F$.
Spectral information about this operator is most conveniently extracted
from the traces ${\rm tr}\,F^n$, where $n$ plays the role of discretised time. 
The analogue of Gutzwiller's trace formula
has been derived by Tabor \cite{sempr}, who found the relation
\begin{equation}
\label{eq1}
{\rm tr}F^n=\sum_{\rm p.\,o.}^{n=rn_0}\frac{n_0}{
\left| 2-{\rm tr}\,M \right|^{1/2}} 
\exp\left({\rm i}JS-{\rm i}\mu\frac{\pi}{2}\right)
\; ,  \end{equation}
between the traces and the periodic orbits of the corresponding
chaotic classical map.
For convenience we denote here the inverse Planck's constant by $\hbar^{-1}=J$.
The sum is made up of all orbits of
primitive (first return) period $n_0$ with $n=n_0r$, $r$ an integer.
The $r$th return of an orbit is characterised by 
the action $S=rS_0$,
trace of the monodromy matrix (linearised map) $M=M_0^r$, and 
the Maslov index $\mu=r\mu_0$ (which for elliptic orbits satisfies
a slightly more involved
composition law under repetitions).

Eq.\ (\ref{eq1}) is valid for completely chaotic systems.
For hyperbolic orbits, the  eigenvalues  of the monodromy matrix $M_0$
are $e^{\pm\lambda_0}$, while elliptic orbits have unimodular eigenvalues
$e^{\pm {\rm i}\omega_0}$.
In both cases the 
expressions for the semiclassical amplitudes 
$A\propto \left| 2-{\rm tr}\,M \right|^{-1/2}$
given in the introduction
follow immediately.
As advertised,  the semiclassical amplitude $A$
of an individual orbit diverges when
$\omega_0=2\pi n/m$, with $n$, $m$ integers (taken relatively prime).
The type of bifurcation depends on $m$, with $m=1$ the tangent
bifurcation, $m=2$ the period-doubling bifurcation,  $m=3$ the
period-tripling bifurcation, and so forth.
Close to a bifurcation one should replace contributions of individual orbits 
by collective contributions in the trace formula, of the form
\begin{equation}
A=\frac{J}{2\pi}
\int_0^\infty{\rm d}I\int_0^{2\pi}{\rm d}\phi\,\Psi(I,\phi)
\exp[{\rm i}J\Phi(I,\phi)],
\label{eq:collective}
\end{equation}
with the `amplitude function' $\Psi$ and the `phase function' $\Phi$
both
depending on the type of bifurcation under consideration.
Here we have used canonical polar coordinates $I$, $\phi$, which
parametrise the phase space of the classical map as
\begin{equation}
p=\sqrt{2I}\sin\phi,\qquad
q=\sqrt{2I}\cos\phi,
\end{equation}
giving for the differentials ${\rm d}p\,{\rm d}q={\rm d}I\,{\rm d}\phi$.
The phase function is a local approximation to the generating 
functions $S(q',p)$ of the classical map $(q,p)\rightarrow (q',p')$.
Right at the bifurcation the amplitude function reduces to $\Psi=1$,
while the phase function is given by simple normal forms.
For generic bifurcations we have
\cite{ozorio,unif123,pver4}
\begin{equation}
\begin{array}{lr}
\Phi=S_0
-\varepsilon q-aq^3-b p^2 &\qquad(m=1),\\
\Phi=S_0-\varepsilon q^2-aq^4-b p^2 &  \qquad(m=2),\\
\Phi=S_0-\varepsilon I-aI^{3/2}\cos 3\phi & \qquad(m=3),\\
\Phi=S_0-\varepsilon I -a I^2 -b I^2\cos 4\phi & \qquad(m=4).
\end{array}
\end{equation}
Here $\varepsilon$ is the bifurcation parameter (bifurcations take place at 
$\varepsilon=0$), while $S_0$, $a$ and $b$ can be regarded as constants.
At the bifurcation ($\varepsilon=0$)
we can rescale the integration
variables $q$, $p$ for $m=1,2$ or $I$ for $m\ge 3$, such that
the combination $J\Phi$ appearing in the exponent of
Eq.\ (\ref{eq:collective})
becomes independent of $J$. What remains is a $J$-dependent prefactor in
front of a $J$-independent integral. We find
$A\propto J^\nu$ with \cite{unif123}
\begin{equation}
\begin{array}{ll}
\nu=1/6 \quad (m=1), \qquad & \nu=1/4 \quad (m=2), \\
\nu=1/3 \quad (m=3), \qquad & \nu=1/2 \quad (m\ge 4) .
\end{array}
\end{equation}
For $\varepsilon\neq 0$ the integral remains $J$-independent
if one also rescales the bifurcation parameter
according to
$\varepsilon'= \varepsilon J^\mu$  \cite{Berryneu},
with 
\begin{equation}
\begin{array}{ll}
\mu=2/3 \quad (m=1), \qquad & \mu=1/2 \quad (m=2),\\
\mu=1/3 \quad (m=3), \qquad & \mu=1/2 \quad (m\ge 4) .
\end{array}
\label{eq:mu}
\end{equation}
These exponents determine the semiclassical range
of the bifurcations in parameter space.

The case $m=3$ is special in the sense that period-tripling bifurcations
are usually accompanied 
by a tangent bifurcation, so 
close in parameter space that the
semiclassical contribution given above 
looses validity for accessible values of $J$. 
This gives the unique opportunity to test also predictions for a
bifurcation of higher codimension. 
The normal form is \cite{pver3,kodim2}
\begin{equation}
\Phi(I,\phi')=S_0-\varepsilon I-aI^{3/2}\cos 3\phi-bI^2
\: .\end{equation}
The tangent bifurcation takes place at $\varepsilon=\frac{9\,a^2}{32\,b}$,
while the period-tripling
bifurcation occurs at $\varepsilon=0$. For $\varepsilon=a=0$
one has to consider an integral of the form
\begin{equation}
J\int_0^\infty{\rm d}I\int_0^{2\pi}{\rm d}\phi\, \exp[{\rm i}JI^2]\propto
J^{1/2},
\end{equation}
and obtains $\nu=1/2$. For $\varepsilon$, $a\neq 0$ we obtain the two
scaling parameters $\mu_\varepsilon=1/2$ and $\mu_a=1/4$ that
characterise the semiclassical range of the bifurcation
in parameter space.

\section{Numerical results}
\label{sec3}

\begin{figure*}
\epsfxsize=0.85\textwidth
\centerline{\epsffile{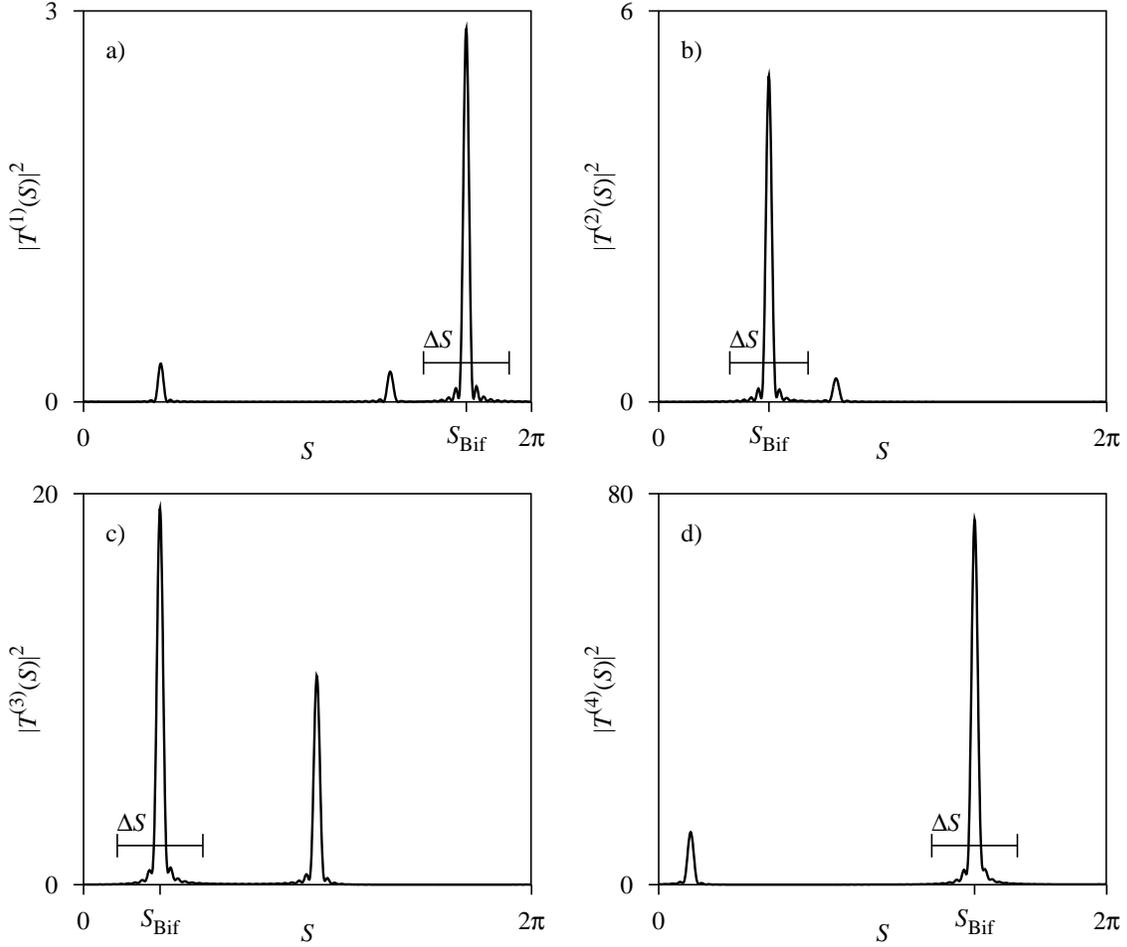}}
\medskip
\caption{The action spectrum $|T^{(n)}(S)|^2$ for the kicked top
close to different types of bifurcations.
For the inverse Fourier transformation (\protect\ref{Bbet1sep})
we restrict the $S$-integration to the intervals of width $\Delta S$
around the centres of the peaks $S_{\rm Bif}$,
eliminating in this way the contributions from 
other periodic orbits.
a) $n=1$,  $k=2.6$. The large peak at $S_{\rm Bif}$ 
comes from orbits which are involved in a tangent bifurcation at
$k\simeq 2.5$. 
b) $n=2$,  $k=2.3$, $\alpha_1=1.39$,
close to the period-doubling bifurcation at $k\simeq 2.1$.
c)
$n=3$,  $k=2$, close to 
two period-tripling bifurcations
at $k\simeq 1.85$ (left peak) and $k\simeq1.97$ (right peak).
d)
$n=4$,  $k=1.2$.
The large peak arises from 
orbits involved in a period-quadrupling bifurcation
at $k\simeq1.0$. (The orbits of the smaller peak 
are also  involved in a period-quadrupling bifurcation, at
$k\simeq1.2$.)}
\label{ft1}
\end{figure*}

\begin{figure}
\epsfxsize=0.5\textwidth
\centerline{\epsffile{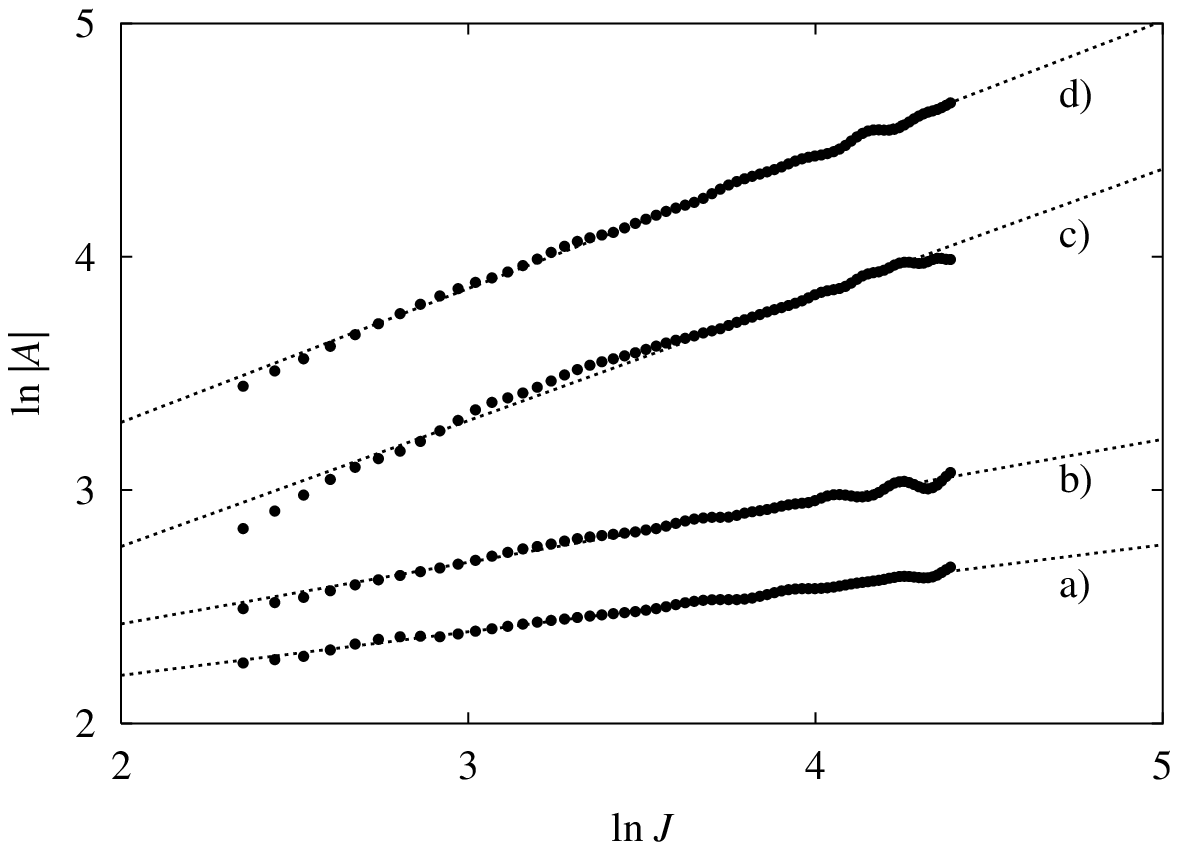}}
%\begin{figure*}
%\epsfxsize=0.9\textwidth
%\centerline{\epsffile{fig2.ps}}
\caption{Logarithmic plots versus inverse Planck's constant $J$
of the maximal (in parameter space) contributions $|A|$
of orbits involved in the bifurcations of Fig.\ \protect\ref{ft1},
calculated by Eq.\ (\protect\ref{Bbet1sep}).
a) Tangent bifurcation.
The average (dotted line) gives the exponent $\nu\simeq0.1866$.
b) Period-doubling bifurcation, $\nu\simeq0.2636$.
c) Period-tripling bifurcation of higher codimension,
$\nu\simeq0.5327$.
d)
Period-quadrupling bifurcation, $\nu\simeq0.5734$.}
\label{bifj1}
%\end{figure*}
\end{figure}

We now wish to investigate how the semiclassical singularities at
bifurcations emerge for a representative dynamical system,
the periodically kicked top \cite{klassik}, which has proven useful in
testing semiclassical results before. 
The dynamics consists of a sequence of rotations and torsions, with
Floquet operator
\begin{eqnarray}
F&=&\exp \left( -{\rm i}\frac{k_1}{2j+1}\hat{J}_z^2-{\rm i}\alpha_1\hat{J}_z
\right)
  \exp \left( -{\rm i}\beta\hat{J}_y \right) 
	\nonumber
	\\
	&&{}\times
  \exp \left( -{\rm i}\frac{k_2}{2j+1}\hat{J}_x^2-{\rm i}\alpha_2\hat{J}_x
	\right)
\; . 
\end{eqnarray} 
The angular momentum operators $\hat{J}_{x,y,z}$ obey the commutator relation 
$[\hat{J}_i,\hat{J}_j]=i\varepsilon_{ijk}\hat{J}_k$. 
Since the square of the angular momentum ${\bf J}^2=j(j+1)$ is conserved the
phase space is the unit sphere.
The role of the inverse Planck's constant
is played by $J=j+\frac{1}{2}$,
which is equal to one half of the Hilbert space
dimension. The semiclassical limit is reached by sending $J\rightarrow\infty$.
We fix the rotation parameters $\alpha_1=0.8$, $\beta=1$, $\alpha_2=0.3$,
and use the torsion strengths $k_1\equiv k$ and $k_2=k/10$ to control the
degree of chaos of the classical map. The system is integrable for $k=0$ and
displays well-developed chaos from $k\simeq 5$. 

The quantum-mechanical evaluation of $F$ is described in Ref.\ \cite{klassik}.
We computed the  traces of the Floquet operator 
and separated the contributions of different (clusters of) orbits by 
evaluating the action spectrum
(a Fourier transformation of the trace with respect to the inverse Planck's
constant)
\cite{actionsp},
\begin{equation}
 T^{(n)}(S)=\frac{1}{j_{\rm max}-j_{\rm min}+1}
 \sum^{j_{\rm max}}_{j=j_{\rm min}} 
            {\rm tr} F^n (j) e^{-{\rm i}(j+\frac{1}{2})S}
\label{aspek}\; , 
\end{equation} 
where the difference $j_{\rm max}-j_{\rm min}$ determines the resolution in $S$
($j_{\rm min}=1$, $j_{\rm max}=100$). 
The results for parameters close to different types of bifurcations are
shown in Fig.\ \ref{ft1}.
The contribution at given $j$ of orbits pertaining to a given peak 
can be obtained by an inverse Fourier transformation,
\begin{eqnarray}
 A &\propto&
 \!\!\!\!\!\int\limits^{S_{\rm Bif}+\frac{\Delta S}{2}}_{S_{\rm Bif}
-\frac{\Delta S}{2}}
\!\!\! {\rm d} S \left( \sum^{j_{\rm max}}_{j'=j_{\rm min}}\!\!
{\rm tr}\, F^n (j') 
e^{-{\rm i}(j'+\frac{1}{2})S} \right) e^{{\rm i}(j+\frac{1}{2})S} 
\nonumber\\ \!\!\!
& = &2\sum^{j_{\rm max}}_{j'=j_{\rm min}} {\rm tr}\, F^n (j')\,
e^{{\rm i}(j-j')S_{\rm Bif}}
   \, \frac{\sin \left[ (j-j')\Delta S/2 \right]}{j-j'}
\label{Bbet1sep}\, ,
\nonumber\\
\end{eqnarray}
where the integral over actions $S$ is restricted to an interval $\Delta S$ 
around the centre $S_{\rm Bif}$ of the peak.
This eliminates contributions of other periodic orbits. 

It is convenient to
tune the control parameter $k$ slightly away from the bifurcation
to the value that maximises the contribution of the bifurcating orbits
to ${\rm tr}\,F^n$.
The parameter $k$ of the maximum approaches
the true bifurcation point with the exponent $\mu$, Eq.\ (\ref{eq:mu}).
The maximal contribution
is of the same order of magnitude as the contribution at the bifurcation,
but it is less sensitive to changes in the parameters.
Most importantly, this procedure does not require any classical information
(like the precise parameter value of the bifurcation) and is
hence genuinely quantum mechanical.
We extract the  exponents  $\nu$ from logarithmic plots of  the maximal
$|A|$ versus $J$, shown in Fig.\ \ref{bifj1}.
In all cases we find good agreement with the theoretical predictions.
For a tangent bifurcation at $k \simeq 2.5$
the observed exponent is $\nu \simeq 0.1866$ (theoretically, $\nu=1/6$).
Two different period-doubling bifurcations appear at $k\simeq 2.8$ and 
produce overlapping peaks in the action spectrum. We separated them by changing
$\alpha_1$ to $\alpha_1=1.39$, moving in that way
one of the period-doubling bifurcations to $k\simeq 2.1$. 
The exponent for this bifurcation is $\nu\simeq0.2636$
(theoretically, $\nu=1/4$).
Back to the original value $\alpha_1=0.8$, 
we find for the period-quadrupling bifurcation at $k\simeq 1.0$ 
the exponent $\nu\simeq 0.5734$ (theoretically, $\nu=1/2$).

As mentioned above, for all known dynamical systems
period-tripling bifurcations are typically accompanied by a tangent 
bifurcation, so close in parameter space that one has to treat
the situation as a bifurcation of higher codimension.
For the kicked top, 
an angular momentum of about $J\simeq 10^5$ would be needed 
for separating the orbits in the
`period-tripling+tangent' bifurcation at $k\simeq 1.85$.
The same is true for a similar sequence of bifurcations at $k\simeq 1.97$.
For the much smaller values of $J$ that we use here
we hence have the unique opportunity 
to test the exponent for a case of higher codimension.
As before,
the result $\nu\simeq 0.5327$
(for the bifurcations at $k\simeq 1.85$)
is close to the theoretical expectation $\nu=1/2$.

\section{Conclusions}
\label{sec4}

We have studied the asymptotic behaviour for $\hbar\to 0$
of periodic-orbit contributions to semiclassical trace formulae,
around points in parameter space where orbits bifurcate.
For the most common types of bifurcations the theoretically predicted
power-law divergence $\propto \hbar^{-\nu}$ was tested numerically
for a representative dynamical system, the
kicked top, giving good agreement for the exponents $\nu$.

In the semiclassical limit
the contribution of non-bi\-furcat\-ing orbits reaches a constant value
$|A|={\cal O}(\hbar^0)$,
corresponding to $\nu=0$. 
It follows from Eq.\ (\ref{eq:collective}) that the exponent for
bifurcating orbits falls into the range $0<\nu<1$.
As a consequence, the semiclassical contribution of bifurcating orbits
is dominant when parameters are close enough to the bifurcation point.
On first sight this seems to require a careful tuning of the parameters.
From the perspective of spectral statistics, however, a careful tuning 
often turns out to be unnecessary \cite{jab,Berryneu}:
The period $n$ of orbits that contribute to the
spectrum increases in the semiclassical
limit as well, and one enters a competition
between the weight of bifurcations in parameter space (given by the
exponents $\nu$ and $\mu$) and the proliferation of their number
with increasing $n$.
Some quantities are dominated by bifurcating orbits even when the
proliferation is not taken into account.
The final outcome of this competition is not clear
at the moment and certainly deserves further investigation.

\begin{acknowledgement}
This work was supported by the Sonderforschungsbereich 237
of the Deutsche Forschungsgemeinschaft and the
Dutch Science Foundation NWO/FOM.
\end{acknowledgement}

\end{document}